\documentclass{aa} 
\usepackage{color}
\usepackage{graphicx}
\usepackage[fleqn]{amsmath}	
\usepackage{amssymb}	
\usepackage{mathtools}
\usepackage{txfonts}
\usepackage{hyperref}
\hypersetup{
  colorlinks=true,   
  citecolor=blue,    
  urlcolor=blue,     
  linkcolor=blue,     
}

\DeclareRobustCommand{\ion}[2]{%
\relax\ifmmode
\ifx\testbx\f@series
{\mathbf{#1\,\mathsc{#2}}}\else
{\mathrm{#1\,\mathsc{#2}}}\fi
\else\textup{#1\,{\mdseries\textsc{#2}}}%
\fi}

\newcommand{\ha}{H$\alpha$}
\newcommand{\hb}{H$\beta$}

\def \arcsec    {$^{\prime\prime}$}

\newcommand{\swift}{\emph{Swift}\,}

\newcommand{\rev}[1]{{ #1}}
\newcommand{\revs}[1]{{ #1}}

\begin{document}

   \title{Low-Eddington ratio, changing-look active galactic nuclei: the case of NGC 4614}


   \author{Elisabeta Lusso\thanks{\email{elisabeta.lusso@unifi.it}}
          \inst{1,2}
          \and
           Lapo Casetti
          \inst{1,2,3}
          \and
          Marco Romoli
          \inst{1,2}
          \and
          Lara Fossi
          \inst{1}      
          \and
          Emanuele Nardini
          \inst{2}
          \and
          Emanuele Arra
          \inst{1}      
          \and
          Benedetta Barsi 
          \inst{4}      
          \and
          Clarissa Calamai 
          \inst{1}      
          \and
          Francesca Campani 
          \inst{4}      
          \and
          Riccardo Capogrosso
          \inst{1}      
          \and
          Francesco Chiti Tegli 
          \inst{1}      
          \and
          Riccardo Ciantini 
          \inst{1}      
          \and
          Eirini Demertzi
          \inst{5}      
          \and
          Marina A. Gaitani
          \inst{6}      
          \and
          Asia Giudice
          \inst{1}      
          \and
          Alessia Gori
          \inst{1}      
          \and
          Lorenzo Graziani
          \inst{1}      
          \and
          Laura Macchiarini
          \inst{1}      
          \and
          Marianna Michelagnoli 
          \inst{1,2}      
          \and
          Chiara Niccolai
          \inst{1}      
          \and
          Irene Parenti
          \inst{1}      
          \and
          Simone Pistolesi
          \inst{1}      
          \and
          Martina Rago
          \inst{1}      
          \and
          Ofelia Romani
          \inst{1}      
          \and
          Leonardo Sani
          \inst{1}      
          \and
          Jacopo Sartini
          \inst{1}      
          \and
          Matilde Scianni
          \inst{1}      
          \and
          Alba Triggianese
          \inst{1}      
          \and
          Gloria Andreuzzi
          \inst{7,8}
          \and
          Filippo Ambrosino
          \inst{8}      
          }

   \institute{Dipartimento di Fisica e Astronomia, Universit\`{a} di Firenze, via G.\ Sansone, 1, 50019 Sesto Fiorentino, FI, Italy
         \and
             INAF-Osservatorio Astrofisico di Arcetri, Largo E.\ Fermi 5, 50125 Firenze, Italy
         \and
             INFN, Sezione di Firenze, via G.\ Sansone, 1, 50019 Sesto Fiorentino, FI, Italy
        \and
             Niels Bohr International Academy, Niels Bohr Institute, Blegdamsvej 17, DK-2100 Copenhagen {\O}, Denmark
        \and
             \'{E}cole Centrale de Lyon, 36 Av.\ Guy de Collongue, F-69130 \'{E}cully, Lyon, France
        \and
             Department of Physics, Aristotle University of Thessaloniki, 54124 Thessaloniki, SKG, Greece
        \and
             Fundaci\'{o}n Galileo Galilei, Rambla Jos\'{e} Ana Fernandez P\'{e}rez 7, 38712 Bre\~{n}a Baja, TF, Spain
        \and
             INAF-Osservatorio Astrofisico di Roma, via Frascati 33, 00078 Monte Porzio Catone, RM, Italy
          }

   \date{\today}

 
  \abstract{
  Active galactic nuclei (AGN) are known to be variable sources across the entire electromagnetic spectrum, in particular at optical/ultraviolet and X-ray energies. Over the past decades, a growing number of AGN have displayed type transitions: 
  \rev{from type 1 to type 2 or viceversa} within a few years or even several months. These galaxies have been commonly referred to as changing-look AGN (CLAGN). Here we report on a new CLAGN, NGC 4614, which transitioned from a \rev{type 1.9} to a type 2 state. \rev{NGC 4614 is a nearly face-on barred galaxy at redshift $z = 0.016$, classified as a low-luminosity AGN.} 
  Its central black hole has a mass of about $1.6\times 10^7\,M_\odot$ and an Eddington ratio around 1 percent. We recently acquired optical spectra of NGC 4614 at the Telescopio Nazionale Galileo and the data clearly suggest that the broad H$\alpha$ component has strongly dimmed, 
  if not disappeared. A very recent \swift\ observation confirmed our current optical data, with the AGN weakened by almost a factor of 10 with respect to previous X-ray observations. Indeed, NGC 4614 had been also observed by \swift/XRT 6 times in 2011, when the source was clearly detected \rev{in all observations}. 
  \rev{By fitting the stack of the 2011 \swift\ observations we obtain a photon index of $\Gamma=1.3\pm0.3$ and an equivalent hydrogen column density of $N_{\rm H}$\,=\,$1.2\pm0.3$\,$\times$\,10$^{22}$ cm$^{-2}$, indicating that NGC 4614 can be moderately absorbed in the X-rays.}
  Although a significant change in the foreground gas absorption that may have obscured the broad line region cannot be entirely ruled out, the most likely explanation for our optical and X-ray data is that NGC 4614 is experiencing \rev{a change in the accretion state that reduces the radiative efficiency of the X-ray corona.} 
  }
  

\titlerunning{Low-Eddington ratio, changing-look AGN (NGC 4614)}
\authorrunning{E.\ Lusso et al.}
\maketitle
%

\section{Introduction}
Active galactic nuclei (AGN) are powered by the accretion of matter onto the supermassive black hole (SMBH; 10$^6-$10$^{10}$ \(M_\odot\)) located at their centre and represent a key stage of the life cycle of galaxies. This active phase is \rev{typically characterised by a (nuclear) luminosity $\gtrsim$\,10$^{43}$ erg\,s$^{-1}$} and significant variability, mostly at X-ray energies. 
Depending on the presence or not of broad emission lines (\rev{with a full-width at half-maximum $>$\,1000} km\,s$^{-1}$) in their optical/UV spectra, AGN are commonly classified into \rev{two broad categories:} type 1 and type 2 AGN, respectively. These two classes can also be interpreted as the same kind of objects surrounded by a dusty and gaseous torus differently oriented towards the observer, which shields the broad-line region (BLR) emission in type 2 but not in type 1 AGN \citep{1993ARA&A..31..473A, 2015ARA&A..53..365N}. \rev{Whilst the fundamental parameters that determine the presence of a BLR are still uncertain, they are likely related to the SMBH accretion state and its mass \citep[see e.g.][]{nicastro2000,Elitzur2009,ch2011,Chakravorty2014}.
Intermediate AGN classifications \revs{\citep[see e.g.][]{Winkler1992,jana2025,Osterbrock1977}} are based on the flux ratio between broad (H$\beta$) and narrow ([\ion{O}{iii}] 5007\AA) lines (types 1.2, 1.5, 1.8), or on the detection of a broad H$\alpha$ with missing H$\beta$ (type 1.9). This finer classification scheme can result} from the combination of viewing angle and a clumpy distribution of the obscuring material of different optical depths; \rev{extinction and AGN/galaxy contrast might also have a role} \citep{1989ApJ...340..190G,BG2024}. 
\rev{Such a classification is based on the spectral properties at optical wavelengths, and is generally in good agreement with the level of obscuration determined at X-ray energies. Specifically, highly obscured (Compton-thick) AGN are characterised by a hydrogen column density $N_{\rm H}$ larger than $10^{24}$ cm$^{-2}$, while a source is considered unobscured if $N_{\rm H} <10^{21}$ cm$^{-2}$. In the intermediate (Compton-thin) range, the optical and X-ray classifications can significantly depart from each other (e.g. \citealt{Koss17,Shimizu18}).}

In the past decades, a number of AGN have been discovered to undergo type transitions within a few years or even several months. These galaxies have been dubbed as changing-look AGN (CLAGN). \rev{CLAGN have been observed at both X-ray energies, transforming from a Compton-thick to a Compton-thin state or vice versa \rev{(e.g. \citealt{2003MNRAS.342..422M,Risaliti05,Bianchi09})}, and at optical wavelengths}, with the broad line component of \ha\ and \hb\ appearing in type 2 AGN or vanishing in previously known type 1 AGN (e.g. \citealt{denney2014, Shappee2014, 2020ApJ...890L..29A, 2020ApJ...901....1W}).
\rev{In the X-ray case, the observed changes in the continuum are usually due to absorption effects, where the obscuring material is located in the very inner regions (sub-parsec, down to a few tens of gravitational radii). In the optical case, the type transition could be also related to a change in the accretion state \citep[e.g.][]{jana2025}. In this sense, optical CLAGN are definitely more intriguing in the context of accretion physics.}

This phenomenon is not limited to local Seyferts. Indeed, the first quasar observed to transition from a type 1 to a type 1.9 (in a period of 10 years) was the X-ray selected galaxy SDSS J015957.64$+$003310.5 (J0159$+$0033) at $z = 0.31$ \citep[][]{lamassa2015}. The changing state for this object was interpreted to be caused by a fading of the AGN continuum, 
reducing the supply of ionising photons available to excite the gas in the immediate vicinity of the SMBH, rather than by variable absorption. 
Vanishing broad-line components have also been observed in NGC 7603 \citep{to1976}, NGC 4151 \citep{pp1984}, NGC 1566 \citep{Alloin1985}, Mrk 372 \citep{Gregory1991}, Mrk 993 \citep{tran1992}, and 3C 390.3 \citep{pp1984,vz1991}, for instance. Broad-line components have instead appeared in Mrk 6 \citep{kw1971}, Mrk 1018 \citep{Cohen1986}, NGC 1097 \citep{storchiberg1993}, and NGC 7582 \citep{Aretxaga1999}. 
These variations (sometimes extremely strong, see \citealt{Shappee2014} for NGC 2617) provide us with key information about the structure of the BLR (at sub-parsec scale) and the intensity (and shape) of the ionising continuum. 
Changes in the AGN continuum, or emission line strengths, or both, could have different physical origins, such as the existence of an obscuring material that covers (totally or partially) the BLR \citep[e.g.][]{elitzur2012}; or changes in the accretion rate \citep[e.g.][]{stern2018} that can be also attributed to episodic accretion events associated with the tidal disruption of stars by the SMBH \citep[e.g. ][]{Eracleous1995}. Yet, in spite of the increasing number of newly discovered CLAGN, the physical mechanism at the origin of CLAGN is still under debate \citep[see][for an extensive review on the topic]{ricci2023}.

Recently, long-slit spectroscopic data acquired by undergraduate students of the University of Florence, Italy, at the Telescopio Nazionale Galileo (TNG) with Dolores in April 2019 unveiled the transition of the Seyfert galaxy NGC 4156 from a type 2 (no broad-line emission) towards a (nearly) type 1, showing a rise in the blue continuum as well as the appearance of a broad component of the \ha\ line, which were both absent in a 2004 SDSS spectrum; follow-up TNG observations performed in 2022 confirmed the presence of the broad \ha\ line, but the blue continuum excess was no longer detected, suggesting that NGC 4156 may be transitioning back to a type 2 \citep{Tozzi2022}. 


Here we report on a new CLAGN, NGC 4614, which transitioned from a type \rev{1.9 (i.e., no broad \hb\ line but presence of a broad component of the \ha\ line) to a type 2} state. Similarly to NGC 4156, NGC 4614 was observed during a campaign led by students at the University of Florence (see Appendix~\ref{IOA} for details). 
NGC 4614 is a nearly face-on barred galaxy at redshift $z = 0.016$, which was \rev{loosely identified as a type 1 by \citet{Oh:2015}} due to the presence of a broad component in the H$\alpha$ emission line. 
The black hole mass for this low-luminosity AGN is $\log M_{\rm BH}/M_\odot\simeq7.2$ \citep[see][for details]{Oh:2015} derived from the line width and luminosity of the broad \ha\ line. The bolometric luminosity is $\log L_{\rm bol}/(\rm erg\,s^{-1})\simeq 43.25$, computed from the relation developed by \citet[][$L_{\rm bol}\simeq3500 L_{[\rm O\,III]}$]{Heckman2004}, implying an Eddington ratio, $\lambda_{\rm Edd}$, of 0.01 ($\lambda_{\rm Edd}=L_{\rm bol}/L_{\rm Edd}$, where $L_{\rm Edd}=1.3\times10^{38}$ erg\,s$^{-1}$). 
The spectrum of this galaxy was acquired at the TNG with Dolores and the LR-B grism on May 8, 2024, in far from optimal conditions due to the presence of clouds. Nonetheless, after reduction, the LR-B spectrum clearly suggested that the broad H$\alpha$ component has strongly dimmed. We then performed higher-resolution follow-up observations of NGC 4614 at the TNG in July 2024, confirming the significant fading, if not the disappearance, of the broad \ha\ component (see Sec.\ \ref{sec:VHR-R}). 
NGC 4614 had been observed in the X-rays by Swift/XRT 6 times in 2011 and the source was clearly detected. 
These were the only X-ray data available for NGC 4614 until we obtained another Swift/XRT snapshot following the discovery of the optical changes, which was performed on November 1, 2024. In the latter data the source was formally undetected (see Sec.\ \ref{Swift data}).

The paper is structured as follows. We present both the optical and X-ray spectroscopic data for NGC 4614 in Secs.~\ref{optobservations} and ~\ref{Swift data}. Section~\ref{Optical spectral analysis} details the spectral fitting procedure employed to compute the spectral properties from the optical spectra.
Finally, we discuss the possible trigger of the changing state of the galaxy in Sect.~\ref{Changing state or variable absorption?}. Conclusions are drawn in Section~\ref{Conclusions}.
Throughout, we adopt a cosmology where $H_0 = 70$ km s$^{-1}$ Mpc$^{-1}$, $\Omega_{\rm M} = 0.3$ and $\Omega_{\Lambda} = 0.7$, giving a luminosity distance of 69.4 Mpc for NGC 4614.

\section{Optical observations}
\label{optobservations}

\subsection{TNG/LR-B data}
\label{sec:LR-B}
\begin{figure}[ht]
\centering
\includegraphics[width=\linewidth]{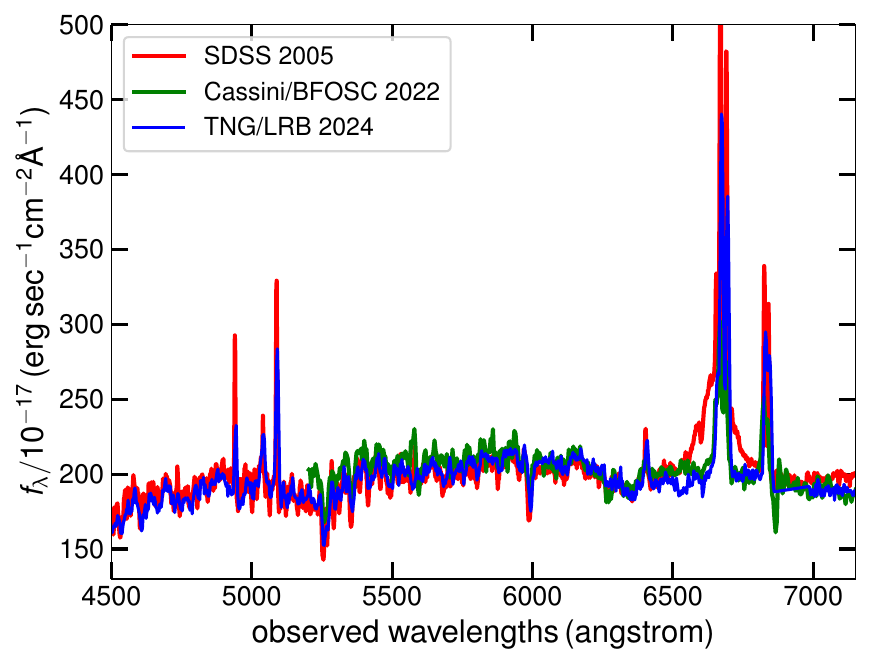}
\caption{Comparison between the archival 2005 SDSS (red curve), the 2022 Cassini/BFOSC (green curve), and the 2024 TNG/LR-B (blue curve) spectra. Fluxes and wavelengths are the observed ones (no Galactic absorption correction has been applied). The absence of the broad H$\alpha$ line is apparent in both the Cassini/BFOSC and the TNG/LR-B data.   \label{fig:spectrum_comparison_full}}
\end{figure}

The spectrum of NGC 4614 was acquired \rev{on May 8, 2024} with the Low Resolution Spectrograph (DoLoRes) mounted at the 3.6m INAF's Telescopio Nazionale Galileo (TNG) using the LR-B grism (1\arcsec\ slit), covering the wavelength range 3000--8400 \AA. The exposure time was 900 s, the airmass of NGC 4614 at the start of the acquisition was 1 with an average seeing of 1.5\arcsec. 
We then reduced the LR-B TNG spectrum with the open source data reduction pipeline \textssc{PypeIt} \citep[][v1.15]{pypeit:joss_pub,pypeit:zenodo}. 
The data reduction follows the standard procedure, including bias subtraction and flat fielding, removal of cosmic rays, sky subtraction, extraction of the 1D spectrum and wavelength calibration. The latter has been derived from the combined helium, neon and mercury arc-lamps in the vacuum frame, and a quadratic polynomial function was used to fit the pixel-wavelength data. The resulting dispersion from the wavelength calibration is $\Delta\lambda=2.74$ \AA/px, with an average $\text{rms}< 0.1$ px. 
A standard star with known spectral type (Feige 66) was observed during the same night to account for telluric absorption and flux calibration, and its data were reduced with the aforementioned pipeline.
The calibrated 1D spectrum of NGC 4614 is shown in Fig.\ \ref{fig:spectrum_comparison_full}, together with the 2005 SDSS spectrum and a spectrum of the same source taken in 2022 at the Cassini telescope in Loiano, Italy (see Sec.\ \ref{sec:Loiano}).

\subsection{TNG/VHR-R data}
\label{sec:VHR-R}
Follow-up long-slit spectroscopic observations of NGC 4614 within the A49DDT3 program (PI: L.\ Casetti) were successfully performed at the TNG during the night of July 4, 2024, in service mode. The weather conditions during the observations were good, with clear sky and a median seeing of 0.7\arcsec. The spectrum of the nuclear region of NGC 4614 was acquired with Dolores and the VHR-R grism, using a slit width of 1\arcsec, in parallactic angle with an integration time of 1800 s. The airmass of NGC 4614 at the start of the observation was 1.3 and the Moon was below the horizon throughout the observations. Besides standard calibrations, the spectrum of the standard A0V star HD109055 was also acquired (same setup as NGC 4614, but for the integration time which was of 45 s) to allow for absolute flux calibration and telluric correction.

We reduced the VHR-R spectrum of NGC 4614 with a custom-made pipeline, using some routines from MAAT (MATLAB Astronomy and Astrophysics Toolbox, \citealt{Ofek:2014}). The data reduction followed the standard procedure, including bias subtraction and flat fielding, removal of cosmic rays, sky subtraction, extraction of the one-dimensional (1D) spectrum, and wavelength calibration. The latter was derived from the Ne-Hg lamp, while a quadratic polynomial function was used to fit the pixel-wavelength data. The resulting dispersion from the wavelength calibration (whose rms error is of 0.25 px) is $\Delta \lambda = 0.78$ \AA/px. The nuclear integrated spectrum was extracted using a pseudo-slit of 1\arcsec $\times$ 5\arcsec centered on the galaxy centre (PA = $60^\circ$, i.e., parallactic angle). The standard star spectrum was reduced with the same pipeline. The telluric correction and the flux calibration were performed with the IDL software XTELLCORR \citep{Vacca_xtellcorr:2003}.

\begin{figure}[ht]
\centering
\includegraphics[width=\linewidth]{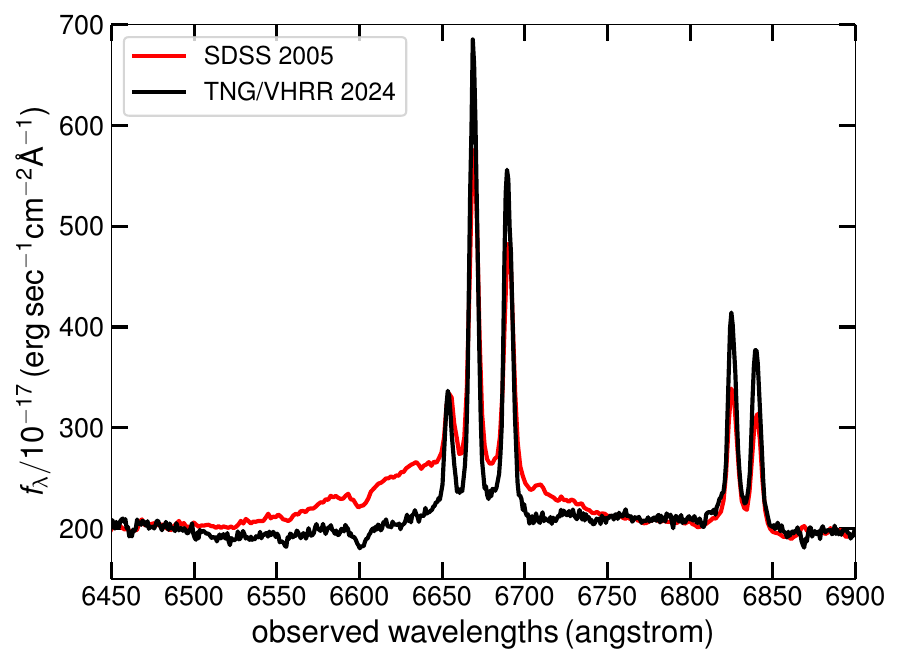}
\caption{Comparison between the archival 2005 SDSS (red curve), and the 2024 TNG/VHR-R (black curve) spectra. Wavelengths are the observed ones, and the spectral region reported in the plot corresponds to that of the H$\alpha$/[\ion{N}{ii}] and [\ion{S}{ii}] line complexes.   \label{fig:spectrum_comparison}}
\end{figure}

The 6450--6900 \AA\ region (in the observed frame) of the calibrated 1D VHR-R spectrum of NGC 4614 is shown in Fig.\ \ref{fig:spectrum_comparison}, together with the 2005 SDSS spectrum. The extraction areas of the two spectra are comparable in size (i.e., roughly 7 arcsec$^2$ for SDSS versus 5 arcsec$^2$ for TNG/VHR-R) although their shapes are different, being the SDSS one circular and ours rectangular. 
The broad component of the H$\alpha$ emission line, clearly visible in the SDSS 2005 spectrum, is apparently no longer detected in the VHR-R spectrum (see Sec.\ \ref{Optical spectral analysis} for a deeper discussion), confirming what suggested by the LR-B data. 

\subsection{Cassini/BFOSC data}
\label{sec:Loiano}
NGC 4614 had also been observed during the night of March 25, 2022 at the Cassini telescope in Loiano, Italy, again during a campaign led by undegraduate students of the University of Florence (see Appendix~\ref{IOA} for details), but the data have been reduced only in 2024, after discovering the change of state of the source. Observations were carried out with BFOSC and grism 4, using a 2\arcsec\ slit and 900 s of integration time; the sky was almost clear and the average seeing was 1.2\arcsec. The acquired data have been reduced using the same tools and procedure used for the TNG/VHR-R data (see Sec.\ \ref{sec:VHR-R}), and the H$\alpha$/[\ion{N}{ii}] and [\ion{S}{ii}] region of the calibrated spectrum is shown in Fig.\ \ref{fig:spectrum_comparison_full} together with the 2005 SDSS archival data and our 2024 TNG/LR-B spectrum. No clear broad contribution to the H$\alpha$ line is visible in the Cassini/BFOSC spectrum, therefore suggesting that NGC 4614 had already changed its state in 2022.  

\section{X-ray observations and analysis}
\label{Swift data}
NGC 4614 was observed by the \swift/X-ray Telescope \citep[XRT;][]{Burrows2005} 6 times in May and July 2011, \rev{with exposure times ranging between 0.7 and 2.7 ks (total exposure 10.9 ks)}. The source was clearly detected in all these observations. XRT data were collected in the photon-counting mode and reduced through the SSDC XRT online analysis service\footnote{\url{https://www.ssdc.asi.it/mmia/index.php?mission=swiftmastr}} (SWXRTDAS; version 3.7.0), which performs data reprocessing and exposure map generation.  
In all the observations, the source spectra were extracted from a circular region of 20-pixel radius located at the source position, corresponding to about 47\arcsec. The background spectra were extracted from an annular region with inner and outer radii of 40 and 80 pixels (94\arcsec\ and 189\arcsec), respectively, around the source extraction area. The quality of the single observations is not sufficient to perform \rev{a detailed analysis of the individual spectra. Yet, we preliminarily fitted each data set with a simple power-law model (subject to Galactic absorption fixed at $N_{\rm H,Gal}$\,=\,1.16\,$\times$\,10$^{20}$ cm$^{-2}$; \citealt{HI4PI}) to determine the observed X-ray luminosity and any variability.} 
We restricted to energies from 0.5 to 8.0 keV for the spectral fitting and we adopted the \citet{Cash1979} statistics, whereby data are binned to at least 1 count bin$^{-1}$. 
We found an average photon index of $\Gamma=0.25\pm0.11$ and a 0.5$-$8 keV band flux $\log\,(F_{\rm X,obs}/{\rm erg\,s^{-1} cm^{-2}})=-11.59\pm0.04$. Variability, a tell-tale signature of AGN, cannot be established at more than the 1$\sigma$ significance. However, despite the \rev{very} unusual spectral shape, the observed X-ray luminosity is far too high \rev{($>$\,10$^{42}$ erg s$^{-1}$ at 2--10 keV)} to be ascribed to the population of galactic X-ray binaries \citep{Mineo2012} or to any ultraluminous X-ray source \citep{Sutton2012}. 

\begin{figure}[ht]
\centering
\includegraphics[width=\linewidth]{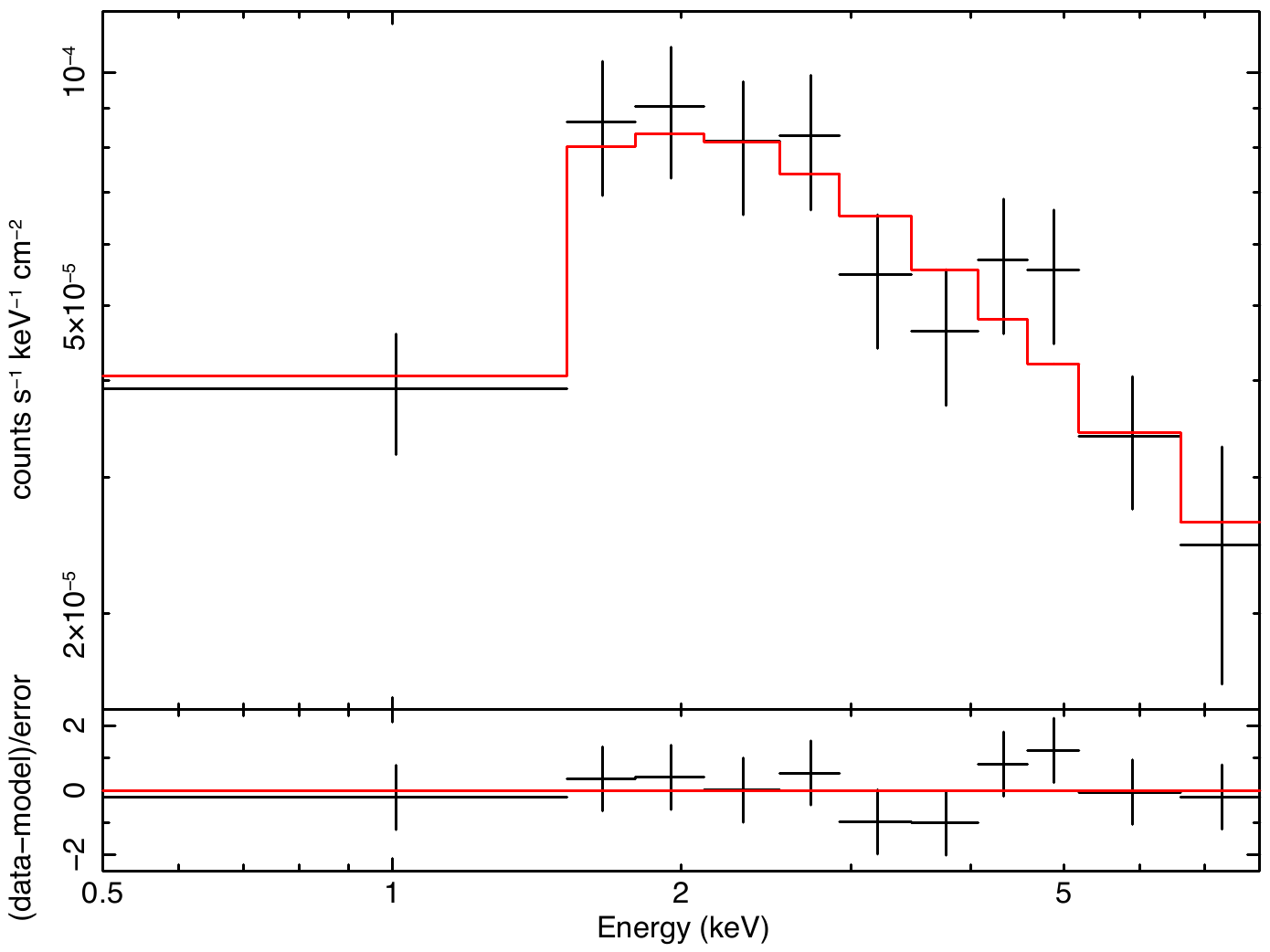}
\caption{\rev{Stacked 2011 \swift\ spectrum and its best fit (solid red line), consisting of an absorbed power law. The data points are binned to a $5\sigma$ significance per energy bin. See Section~\ref{Swift data} for details.} \label{fig:swiftstack}}
\end{figure}

\begin{figure}[ht]
\centering
\includegraphics[width=\linewidth]{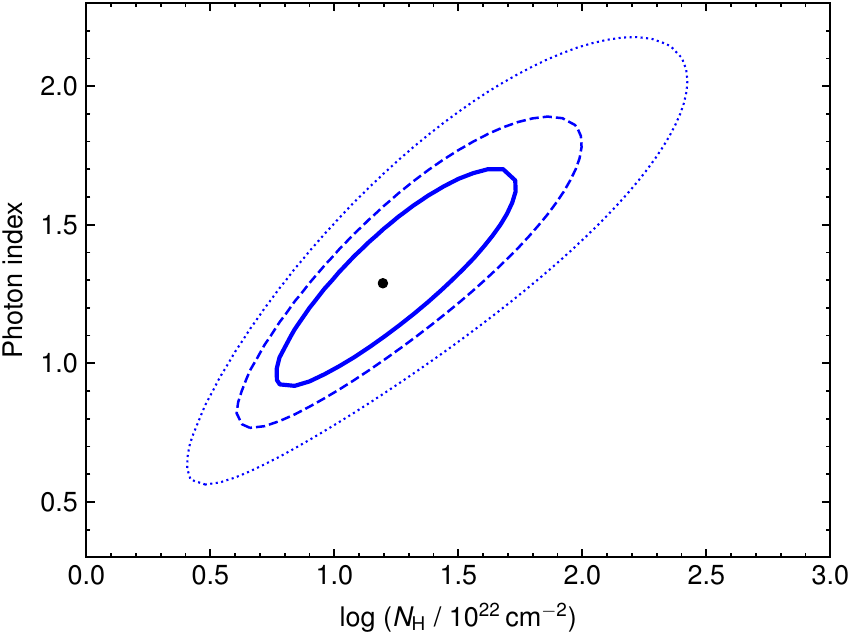}
\caption{Statistical contours \rev{at the 68, 90, and 99 percent level} for the photon index and the equivalent hydrogen column density resulting from the \rev{fit of the stack of the 2011 \swift/XRT spectra,} where the model considers $\Gamma$ and $N_{\rm H}$ free to vary (see Section~\ref{Swift data}). \label{fig:swiftcont}}
\end{figure}

\rev{The previous fit is broadly acceptable ($C/\nu$\,=\,272/242, where $\nu$ is the number of degrees of freedom), but is still not optimal. To gain more insight into the actual spectral shape of NGC 4614, we stacked the six \swift\ spectra to improve the signal-to-noise ratio. With about 280 cumulative net counts in the 0.5$-$8 keV band, we grouped the resulting spectrum to a significance of 5$\sigma$ per energy bin to allow the use of the standard $\chi^2$ minimisation in the fit. A power-law continuum modified by Galactic absorption only does not provide an adequate description of the stacked spectrum, with $\chi^2/\nu$\,=\,31/9 for $\Gamma\simeq0.2$ and a corresponding null-hypothesis probability of 2.5\,$\times$\,10$^{-4}$. The inclusion of an additional absorber in the local frame of NGC 4614 dramatically improves the fit down to $\chi^2/\nu$\,=\,5/8, returning a photon index of $\Gamma=1.3\pm0.3$ and an equivalent hydrogen column density of $N_{\rm H}$\,=\,$1.2\pm0.3$\,$\times$\,10$^{22}$ cm$^{-2}$. The stacked 2011 \swift\ spectrum and its best fit are shown in Figure~\ref{fig:swiftstack}, while the confidence contours for the two parameters of interest in this model are shown in Figure~\ref{fig:swiftcont}. The continuum slope is still somewhat flat, yet fixing it to the more canonical value of 1.8 still leads to a fully acceptable fit ($\chi^2/\nu$\,=\,8/9) with an increase of the absorbing column by $\sim$50 percent. Since the implied X-ray obscuration is moderate, the exact values of $\Gamma$ and $N_{\rm H}$ have a marginal effect on the estimated intrinsic 2--10 keV flux and luminosity, which are $2.7\pm0.3$\,$\times$\,10$^{-12}$ erg\,s$^{-1}$ cm$^{-2}$ and $1.55\pm0.15$\,$\times$\,10$^{42}$ erg\,s$^{-1}$, respectively, from the nominal best fit.
}

We followed up the galaxy with \swift\ as a target of opportunity (ToO, programme ID=16856, PI: E. Lusso) on November 1, 2024 for 1597 s, and the galaxy is formally not detected (significance \rev{$<$\,2$\sigma$ at 0.5$-$8 keV}). We derived a \rev{95 percent} upper limit on the source counts in a 47\arcsec\ (20-pixel) aperture circle of $\lesssim$10 counts. This corresponds to a \rev{2$-$10 keV flux of 4.5\,$\times$\,10$^{-13}$ erg\,s$^{-1}$ cm$^{-2}$, which is} at least 6 times lower than the average of 2011 assuming the same spectral shape. This observation is in agreement with the optical spectra that do not show any sign of nuclear emission from the AGN.
Since there are no other deeper X-ray data for this galaxy, we cannot further test whether absorption or variability are present. \rev{However, we note that, for such a flux drop to be entirely due to a change in the X-ray obscuration, the column density should have increased by at least a factor of $\sim$50.} 

Interestingly, an extended source detected in the first six months of the eROSITA sky scan (eRASS1, from 11 December 2019 to 11 June 2020; \citealt{merloni2024}), namely \rev{1eRASS J124130.8+260242,} is marginally consistent\footnote{\rev{The estimated source extent is 9.8\arcsec\,$\pm$\,5.7\arcsec, while the distance from NGC 4614 is about 12\arcsec.}} with the coordinates of NGC 4614. 
\rev{The 0.2--2.3 keV flux of this extended source is 1.5\,$\times$\,10$^{-13}$ erg\,s$^{-1}$ cm$^{-2}$, which is smaller by a factor of 1.7 than the observed flux of NGC 4614 in the same band as derived from the 2011 \swift\ stack. While the association with NGC 4614 is dubious, the eROSITA detection can serve as a conservative upper limit. Moreover, it implies that NGC 4614 itself would have been readily detected in eRASS1 as a point-like source, were it found in the same X-ray state of 2011. This is a further suggestion that possibly} the AGN is in a prolonged low state (see Section \ref{sec:Loiano}).


\begin{table*}
	\centering
	\caption{Integrated luminosities and EWs resulting from the fit for the main emission lines. For each line measurement, both the broad (B) and narrow (N) components are provided if constrained by the fit. Gaussian profiles are considered. Luminosities and EWs are in units of 10$^{40}$ erg s$^{-1}$ and \AA, respectively. Measurements marked with $^\ast$ are not well constrained (i.e. have a warning flag in the fit).} 
	\label{tab:spfit}
	\centering
	\begin{tabular}{lcc|ccccccc} 
		\hline
		Year & $L^{\rm B}_{\rm H\alpha}$ & EW$^{\rm B}_{\rm H\alpha}$ & $L^{\rm N}_{\rm H\beta}$ & EW$^{\rm N}_{\rm H\beta}$ & $L^{\rm N}_{\rm H\alpha}$ & EW$^{\rm N}_{\rm H\alpha}$ & $L^{\rm N}_{\rm [OIII]}$ & EW$^{\rm N}_{\rm [OIII]}$\\
		\hline
		2005/SDSS  & $5.74\pm0.06$ & $50.7\pm0.5$ & $0.3\pm0.1$$^\ast$ & $2.4\pm0.1$$^\ast$ & $1.30\pm0.02$ & $11.4\pm0.2$ &   $0.51\pm0.02$ & $4.8\pm0.1$ \\
		2024/LR-B  & - & - & $0.34\pm0.03$ & $3.0\pm0.2$ & $2.0\pm0.1$ & $18.6\pm0.6$ & $0.85\pm0.03$  & $8.1\pm0.3$ \\
		2024/VHR-R & $1.13\pm0.02$ & $9.8\pm0.2$ & - & - & $1.40\pm0.02$ & $12.0\pm0.1$ & - & -\\
		\hline
	\end{tabular}
\end{table*}

\begin{figure}[ht]
\centering
\includegraphics[width=\linewidth]{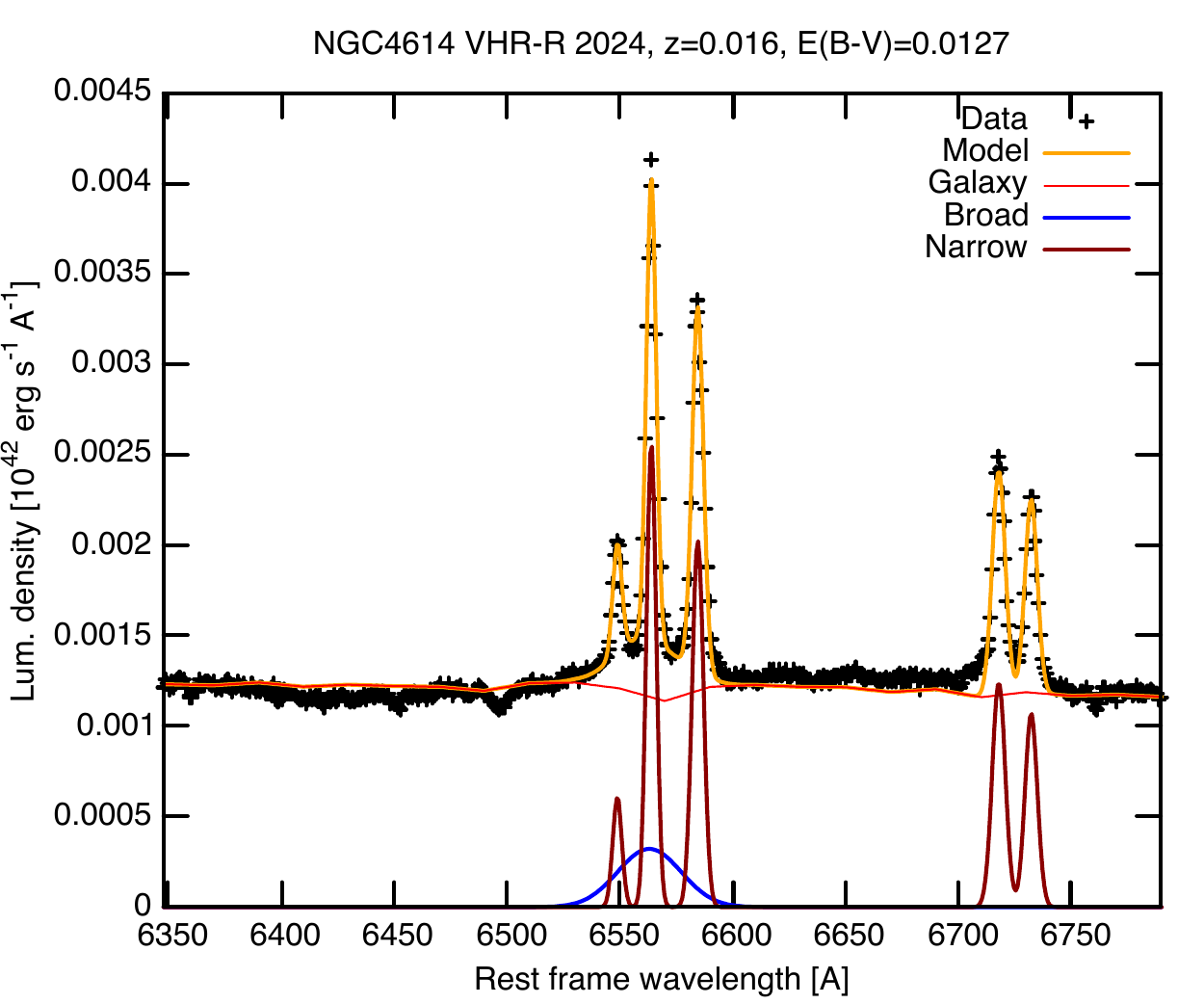}
\caption{ Comparison between the VHR-R spectrum and the overall model, with the individual components shown in the legend. \label{fig:fit_vhrr}}
\end{figure}
\section{Optical spectral analysis} 
\label{Optical spectral analysis} 
We fitted the three 1D spectra of NGC 4614 (i.e. the SDSS 2005, the TNG/LR-B 2024, and the VHR-R 2024) with the fitting code \textssc{QSFit} \citep{2017MNRAS.472.4051C}. Specifically, the SDSS and the LR-B spectra were fitted, assuming Gaussian line profiles, over the 3800--7000 \AA\ wavelength range, while the VHR-R one over the 6200--7800 \AA\ range. 

We first fit the SDSS spectrum, finding that a 13 Gyr old elliptical \citep{Silva1998} provides the best representation of the host galaxy continuum (mostly dominated by the bulge) for NGC 4614, which we then fixed to fit the LR-B and VHR-R data. The [\ion{O}{iii}] 5007\AA\ emission line luminosity measured from the SDSS data is fully consistent with the value reported by \citet[][]{oh2013} (i.e. $5\times10^{39}$ erg\,s$^{-1}$). Given the lower spectral resolution of the LR-B data, the fit of the LR-B spectrum provides slightly higher but consistent emission line luminosity ($\simeq8\times10^{39}$ erg\,s$^{-1}$). This value is also in excellent agreement with the [\ion{O}{iii}] vs 2$-$10 keV correlation \citep[e.g.][]{Ueda2015}, assuming the 2011 X-ray flux state.
From equation (1) in \citet[][]{zhang2022}, we can estimate the NLR size from the [\ion{O}{iii}] line luminosity to be about 1500 pc, consistent with what is found in low-redshift AGN. The light crossing time of a region of this size is a few thousand years, thus no variation on the [\ion{O}{iii}] emission line is expected on the timescales of our observations.

The best-fit parameters for the line luminosity, FWHM, and EW of the broad component detected in the SDSS spectrum are $(5.74\pm0.06)\times10^{40}$ erg\,s$^{-1}$, $6160\pm70$ km\,s$^{-1}$, and $50.7\pm0.5$ \AA, whilst the ones for the narrow component are $(1.30\pm0.02)\times10^{40}$ erg\,s$^{-1}$, $300\pm5$ km\,s$^{-1}$, and $11.4\pm0.2$ \AA, respectively. 
We then tested whether the presence of a broad H$\alpha$ component can be detected in the LR-B and VHR-R data. 
By fitting the VHR-R spectrum, we find that a broad component is barely detected, with the best-fit line luminosity, FWHM, and EW of $\sim$\,$1\times10^{40}$ erg\,s$^{-1}$, $\sim$\,1500 km\,s$^{-1}$, and $\sim$\,10 \AA, respectively; conversely the values of the spectral parameters for the narrow component are fully consistent with the ones computed from the SDSS spectrum. The VHR-R spectrum and the best-fit model are shown in Figure~\ref{fig:fit_vhrr}. The broad component cannot be constrained from the LR-B spectrum.
Moreover, the broad H$\alpha$ is no longer statistically significant in the VHR-R spectrum if a Lorentzian profile is considered instead of a Gaussian, implying that the strength of this feature in 2024 is indeed very low \rev{and that the value given above for the luminosity of the broad H$\alpha$ line can thus be considered as an upper limit}. A summary of the emission line properties is given in Table~\ref{tab:spfit}.

\section{Changing accretion state or variable absorption?}
\label{Changing state or variable absorption?}

\citet{Elitzur2009} noticed that the disc-wind scenario in AGN implies the disappearance of both the BLR and the near-infrared emission from the toroidal obscuration region at luminosities lower than $5\times10^{39}\,(M_{\rm BH}/10^7 M_\odot)^{2/3}$ erg\,s$^{-1}$. The predicted radiative efficiency of accretion onto the SMBH should be lower than $10^{-3}$ for these sources, indicating that their accretion is advection-dominated \citep[][]{Ho2009}. It is unclear whether this scenario can be a viable explanation for the dimming of the broad \ha\ line in NGC 4614, as there is no near/mid-infrared spectroscopy available that covers the range where we expect the nuclear reprocessed emission from dust to be significant (at wavelengths longer than 2\,$\mu$m). The nuclear continuum is also undetected in all the optical spectra, so it is unknown by how much the accretion rate might have changed. The nuclear emission of NGC 4614 is detected at X-ray energies only in 2011, so the AGN should have been active during that period but there is no coeval optical spectroscopy available to constrain any possible variation on the continuum radiation.

If the change in the broad line component of the \ha\ is connected to a drop in the nuclear luminosity, thus in the accretion rate, this should have dropped to values lower than the one observed in 2005, i.e. $\lambda_{\rm Edd} \simeq0.01$. 
\rev{The source is undetected in the latest 2024 \swift\ X-ray observation, yet we can estimate its actual 2$-$10 keV luminosity to be $\approx$\,$1.5\times 10^{41}$ erg s$^{-1}$, i.e. roughly 15 times that of the 2024/VHR-R broad H$\alpha$ line following \citet[][see their Fig.~2]{ho2001}. This is consistent with the inferred upper limit and with the drop of almost one order of magnitude of the X-ray luminosity. Assuming a bolometric correction of 10--15 \citep[][see also \citealt{duras2020}]{lusso2012}, this suggests a current Eddington ratio of $\approx 10^{-3}$.}

The value of $\lambda_{\rm Edd}$ of 1 percent is expected for the accretion disc to transition between a radiatively inefficient accretion flow and a thin accretion disc \citep[e.g.][]{ny1995}, thus suggesting that the low-luminosity (low-Eddington) AGN in NGC 4614 could be operating around the threshold mass accretion rate between these two accretion modes. 
This behaviour is consistent with NGC 4156 and several other local AGN, such as NGC 3516 \citep{Collin-Souffrin1973}, NGC 7603 \citep[e.g.][]{to1976,Kollatschny2000}, Mrk 590 \citep{denney2014}, HE 1136$-$2304 \citep{Parker2016,Zetzl2018,Kollatschny2018}, NGC 2992 \citep{Guolo2021}, and IRAS 23226$-$3843 \citep{Kollatschny2020}. 
Indeed, it has been suggested that all CLAGN could be associated with an accretion state transition at $\lambda_{\rm Edd}$ around 1 percent, similar to what is seen in X-ray binaries \citep[e.g. ][]{nodadone2018,MacLeod2019,hagen2024}, and our recent X-ray observation from \swift\ supports a scenario where the activity of the central AGN has significantly dropped since 2011.
At relatively low accretion rates, the accretion flow is more likely to experience instabilities that drive changes in the optical flux by a factor of a few on multi-year timescales, as also suggested by the most rapidly variable quasars from the SDSS \citep[e.g.][\rev{see also \citealt{jana2025} for recent results from the BAT AGN spectroscopic survey}]{Rumbaugh2018}. 

At present, however, we cannot completely rule out that the observed behaviour is due to obscuration effects.
Following \citet[][]{lamassa2015}, we can estimate the crossing time for an intervening object outside the BLR on a Keplerian orbit to be
\begin{equation}
    t_{\rm cross} = 0.07 \left[ \frac{r_{\rm orb}}{1 \rm{lt-day}}\right]^{3/2} M_8^{-1/2} \arcsin\left[ \frac{R_{\rm BLR}}{r_{\rm orb}}\right]\, {\rm years},
\end{equation}
where $r_{\rm orb}$ is the orbital radius of the foreground object, and $M_8$ is the black hole mass in units of $10^8 M_\odot$. The size of the BLR ($R_{\rm BLR}$) in NGC 4614 relative to the broad H$\alpha$ component, as determined from the SDSS data, is computed from equation (2) in \citet[][with parameters given in the second line of their Table 3]{dallabonta2024} and is 3.7 days (i.e. 0.003 parsecs). If we assume that the intervening object is located at 5 times $R_{\rm BLR}$, $t_{\rm cross}$ is about 3 years. This value increases to $\simeq13$ years in the case we assume $r_{\rm orb}=100\,R_{\rm BLR}$. From these timescales, we cannot fully exclude variable absorption to be the cause of the spectral change in NGC 4614 in the time-frame spanned by the data where the H$\alpha$ is not detected (since 2022). A reappearance of the broad H$\alpha$ component in the next few years might be expected in this ``eclipsing'' scenario.


\section{Conclusions}
\label{Conclusions}

Over the past decades, a growing number of AGN have displayed, sometimes dramatic, \rev{type transitions (or transitions between subtypes)} within a few years or even several months. These galaxies, commonly known as changing-look AGN (CLAGN), are of paramount importance since they could provide key insights on the accretion physics onto the SMBH. Indeed, it is been suggested that many changing-look AGN show spectral variations when they cross the state transition boundary of $\lambda_{\rm Edd}$ of about one to a few per cent.

Here we have reported on new observations of NGC 4614, a low-luminosity AGN \citep[][]{Oh:2015}, which shows a transition from a \rev{type 1.9} to a type 2 state.  
We observed NGC 4614 spectroscopically at the Telescopio Nazionale Galileo in the optical and the data clearly suggest that the broad H$\alpha$ component has at least strongly dimmed, if not disappeared, with respect to archival data taken two decades ago. We have also presented a spectrum of the same source taken at the Cassini telescope in Loiano in 2022, where no broad component of the H$\alpha$ line can be detected either, thus suggesting that the transition had already occurred in 2022. A very recent \swift\ observation is in full agreement with our current optical data, with the AGN weakened by almost a factor 10 with respect to previous X-ray observations. NGC 4614 was observed by the \swift/XRT 6 times in 2011 and the source was clearly detected\rev{, revealing an active AGN. The best fit of the stacked \swift\ data returned a continuum photon index of $\Gamma=1.3\pm0.3$ and a column density of $N_{\rm H}$\,=\,$1.2\pm0.3$\,$\times$\,10$^{22}$ cm$^{-2}$, corresponding to an intrinsic 2$-$10 keV luminosity of $1.55\times10^{42}$ erg s$^{-1}$. The suggested X-ray absorption should be strongly variable (by a factor of at least $\sim$50) to fully account for} the recent X-ray drop. If the obscurer were located beyond the BLR, this could possibly also be associated with the disappearance of the broad H$\alpha$ line.
Yet, the latter scenario requires a substantial degree of geometrical fine-tuning, also considering that NGC 4614 is a nearly face-on system. The most likely explanation for our optical and X-ray data is therefore that NGC 4614 has experienced a change in the accretion state \citep[e.g.][]{nodadone2018}, which likely occurred before 2022, when the Loiano spectrum confirms that the broad H$\alpha$ line was already lacking, and possibly before 2020, when eROSITA visited the source and found it in a low X-ray state compatible with the current one.

This study demonstrates that having complementary X-ray observations during the epoch of the optical/BLR change in look is fundamental to make further progress in understanding the origin of these phenomena. The X-rays, indeed, are the only probe of the properties of the inner accretion flow. So far, no univocal picture has emerged when enough X-ray coverage of optical changing-look events is available. 
NGC 4614 is one of the few sources where X-ray and optical data are almost simultaneous, suggesting an inactive phase of the black hole.
\revs{Further spectroscopic observations, possibly simultaneous, at optical and X-ray energies would be key to better constrain the duty cycle of this galaxy. New optical spectroscopic observations are already planned for the near future.}

\begin{acknowledgements}
We thank the anonymous reviewer for their thorough reading and for useful comments that have improved the clarity of the paper.
Based on observations made with the Italian Telescopio Nazionale Galileo (TNG) operated by the Fundación Galileo Galilei (FGG) of the Istituto Nazionale di Astrofisica (INAF) at the Observatorio del Roque de los Muchachos (La Palma, Canary Islands, Spain) and with the G.\ D.\ Cassini telescope operated by INAF-Osservatorio di Astrofisica e Scienza dello Spazio di Bologna at the Osservatorio di Loiano, Italy. We are grateful to the Swift team for approving the DDT observation of NGC 4614.
We acknowledge the use of public data from the Swift data archive. This research has made use of the XRT Data Analysis Software (XRTDAS) developed under the responsibility of the ASI Science Data Center (ASDC), Italy, and of the NASA/IPAC Extragalactic Database (NED) which is operated by the Jet Propulsion Laboratory, Caltech, under contract with the National Aeronautics and Space Administration. 
This work was performed in part at the Aspen Center for Physics, which is supported by National Science Foundation grant PHY-2210452. EL and EN acknowledge partial support from the Simons Foundation (1161654, Troyer).
    We thank the TNG Director A.\ Ghedina for Director Discretionary Time allowing us to perform the 2024 follow-up higher-resolution observations, for which we also thank W.\ Boschin as support astronomer. We thank I.\ Bruni for the support during the 2022 Cassini observations. The participation of the students to the 2024 TNG observing campaign was financially supported by INAF-Osservatorio Astrofisico di Arcetri and by the Dipartimento di Fisica e Astronomia, Universit\`{a} di Firenze. We thank again the TNG Director and the whole TNG staff for making such an experience possible. 
\end{acknowledgements}   

   \bibliographystyle{aa} 
   \bibliography{NGC4614_paper.bib} 

\begin{appendix}
    \section{Student observing campaign}
    \label{IOA}
    The observations of NGC 4614 carried out at the Cassini telescope in Loiano on March 25, 2022 and at the TNG on May 8, 2024 were part of a multi-year campaign planned and performed by undergraduate students of the Physics and Astronomy Department of the University of Florence (Italy), as part of a course on introductory observational astrophysics \citep{Lusso_did:2022}. 

    This programme targets active and non-active spiral galaxies to perform both imaging and low-to-medium resolution spectroscopic observations with the main scientific aim of measuring the apparent diameter (from imaging) and redshift (from spectroscopy) of these sources to constrain the expansion rate of the universe, parameterised by the Hubble parameter, $H_0$. Ancillary science goals are the determination of the star formation rate and the estimate of the mass of the central supermassive black hole whenever the observed galaxy has an active nucleus. Therefore, any CLAGN observed within this programme is a serendipitous discovery, as no previously known CLAGN has been explicitly targeted. We note that also the discovery of NGC 4156 as a CLAGN  \citep{Tozzi2022} was based on observations carried out in 2019 at the TNG by undergraduate students attending the same course. 
    
    The programme started in 2010. To date, 93 bright nearby ($z < 0.15$) spiral galaxies have been observed using various facilities: 74 at the 1.5m Cassini telescope in Loiano (operated by INAF Bologna, Italy), 4 at the 1.8m Copernico telescope in Asiago-Ekar (INAF Padova, Italy), and 15 at the 3.6m TNG telescope. Each year, data from the observations are reduced by the students as part of their laboratory activity. However, typically not all the sources are reduced soon after the observations. In particular, the 2022 data of NGC 4614 taken at Loiano have been reduced only in 2024, after TNG data for the same source hinted at the disappearance of the broad component of the H$\alpha$ line.
\end{appendix}
\end{document}